\documentclass[aps,prd,showpacs,onecolumn,preprintnumbers,notitlepage,superscriptaddress,
nofootinbib,amsfont,amssymb,10pt,singlespacing] {revtex4-1}
\usepackage{amsmath,bm}
\usepackage{times}
\usepackage{braket}
\usepackage{color,graphicx}
\usepackage{slashed}
\usepackage{hyperref}

\newcommand{\beq}{\begin{eqnarray}}
\newcommand{\eeq}{\end{eqnarray}}
\newcommand{\non}{\nonumber\\ }

\newcommand{\acp}{{\cal A}_{CP}}

\def\lsim{ {\ \lower-1.2pt\vbox{\hbox{\rlap{$<$}\lower6pt\vbox{\hbox{$\sim$}
}}}\ } }
\def\gsim{ {\ \lower-1.2pt\vbox{\hbox{\rlap{$>$}\lower6pt\vbox{\hbox{$\sim$}
}}}\ } }
\definecolor{Red}{rgb}{1.,0.,0.}

\definecolor{Blue}{rgb}{0.,0.,1.}

\definecolor{nicered}{rgb}{0.7,0.1,0.1}
\definecolor{nicegreen}{rgb}{0.1,0.5,0.1}

\hypersetup{colorlinks,citecolor=nicegreen,linkcolor=nicered}

\begin{document}

\title{
Revisiting the $B^{0} \to \pi^{0}\pi^{0} $ decays in the perturbative QCD approach}

\author{Yun-Feng~Li}
\email[Electronic address:]{liyun1990405@163.com}
\affiliation{School of Physical Science and Technology,
 Southwest University, Chongqing 400715, China}

\author{Xian-Qiao~Yu}
\email[Electronic address:]{yuxq@swu.edu.cn}
\affiliation{School of Physical Science and Technology,
Southwest University, Chongqing 400715, China}

\date{\today}

\begin{abstract}
We recalculate the branching ratio and CP asymmetry for $\bar{B}^{0} (B^{0})\to \pi^{0}\pi^{0}$ decays in the Perturbative
QCD approach. In this approach, we consider all the possible diagrams including non-factorizable contributions and annihilation contributions. We obtain $Br(\bar{B}^{0} (B^{0})\to \pi^{0}\pi^{0})=(1.17_{-0.12}^{+0.11})\times 10^{-6}$. Our result is in agreement with the latest measured branching ratio of $B^{0}\to\pi^{0}\pi^{0}$ by the Belle and HFAG
Collaborations. We also predict large direct CP asymmetry and mixing CP asymmetry in $B^{0}\to\pi^{0}\pi^{0}$ decays, which can be tested by
the coming Belle-II experiments.

\end{abstract}

\maketitle

%
%

\section{Introduction}

The detailed study of $B$ meson decays is a key source of testing the Standard Model(SM),
exploring CP violation and in searching of possible new physics beyond the SM.
The theoretical studies of $B$ meson decays have been explored widely in the
literature, especially the nonleptonic two-body branching ratios and their CP asymmetries. Although we have achieved great success in explaining many
 decay branching ratios, there are
still some puzzles remaining. One of the challenges is that the measured branching ratio~\cite{Beringer2012,Amhis,Aubert2003} for the decay of $B$ meson to neutral pion pairs $B^{0}\to\pi^{0}\pi^{0}$ is significantly larger than
the theoretical predictions obtained in the QCD factorization approach(QCDF)~\cite{Beneke2010,Beneke2003,Burrel2006,Beneke2006} or a perturbative QCD approach(PQCD)~\cite{LUY}.

For a long time, the factorization approach (FA)~\cite{MW1987}
was the method we widely use to estimate the decays~\cite{AAli1999, YHChen1999}.
Although the way is an easy method at predictions of branching ratios and in accord with
experiments in most cases, there are still some unclear theoretical points.
In order to study the non-leptonic $B$ decays better,
QCD factorization~\cite{MBeneke1999}and Perturbative QCD approach~\cite{HnLi1995} are
invented. The basic idea of PQCD method is that the transverse momenta
$k_T$ of valence quarks are considered in the calculations of hadronic matrix elements, and then
for $B$ meson decays, non-factorizable spectator
and annihilation contributions are all calculable in the framework of $k_T$ factorization,
where three energy scales $m_{W}$, $m_{B}$, and $t\approx \sqrt{m_B \Lambda_{\rm QCD}} $
are involved~\cite{LUY,HnLi1995,Keum2001}.

The branching ratio
of $B^{0}\to\pi^{0}\pi^{0}$ has been measured, whose data~\cite{Heavy} are
\begin{align}
\left(
\begin{matrix}
         & (1.83\pm 0.21 \pm 0.13)\times 10^{-6}; (BABAR), \\
         & (0.90\pm 0.12 \pm 0.10)\times 10^{-6}; (Belle), \\
         & (1.17\pm 0.13)\times 10^{-6}, (HFAG).
 \end{matrix}\right).
\label{exp1}
\end{align}

In the last more than 10 years, many theoretical teams have calculated this decays in different approach. Beneke and Neubert made the analysis of
 $B^{0}\to\pi^{0}\pi^{0}$ decay based on QCD factorization in 2003~\cite{Beneke2003}. Recently, Qin Chang~\cite{QChang2014}, Xin Liu~\cite{XLiu2015} and Cong-Feng Qiao~\cite{CFQiao2015} {\it et al.} recalculated this decay model using different method. The next-leading-order (NLO) contributions from the vertex
corrections, the quark loops, and the magnetic penguins have also been calculated in the
literature~\cite{SNandi2007,YLShen2011, YMWang2012, SMishima2006}. By comparing their results, we find the agreement between the theoretical
predictions and the experimental data is still not satisfactory, so we revisit the decays of
$B^{0}\to\pi^{0}\pi^{0}$ in this paper. We use the PQCD approach to recalculate this decays directly, non-factorizable contributions and annihilation contribution are all taken into account. Our theoretical formulas about the decay $\bar{B}^{0}\to\pi^{0}\pi^{0}$
in PQCD framework are given in the next section. In Sec.~\ref{sec:numer} we give the numerical
results and discussions of the branching ratio and CP asymmetries. In the end, we give a short summary in Sec.~\ref{sec:summary}.

%
%
\section{ The framework and perturbative Calculations}\label{sec:pert}

For the considered $\bar{B}^{0}\to\pi^{0}\pi^{0}$ decays, the corresponding weak effective Hamiltonian
can be given as ~\cite{G.Buchalla:1996}.
 \beq
 H_{\rm eff}\, &=&\, {G_F\over\sqrt{2}}
\biggl\{ V^*_{ud}V_{ub} [ C_1(\mu)O_1(\mu)
+C_2(\mu)O_2(\mu) ]
 - V^*_{td}V_{tb} [ \sum_{i=3}^{10}C_i(\mu)O_i(\mu) ] \biggr\}+ {\rm H.c.}\;,
\label{eq:exp2}
\eeq
where $G_F$ is the Fermi constant, $C_i(\mu)(i=1,\cdots,10)$ are Wilson coefficients at the renormalization scale
$\mu$ and $O_i(i=1,\cdots,10)$ are four-quark operators
\begin{enumerate}
\item[]{(1) current-current(tree) operators}
\begin{eqnarray}
{\renewcommand\arraystretch{1.5}
\begin{array}{ll}
\displaystyle
O_1\, =\,
(\bar{u}_\alpha u_\alpha)_{V-A}(\bar{d}_\beta b_\beta)_{V-A}\;,
& \displaystyle
O_2\, =\, (\bar{u}_\alpha b_\alpha)_{V-A}(\bar{d}_\beta u_\beta)_{V-A}\;;
\end{array}}
\label{eq:operators-1}
\end{eqnarray}

\item[]{(2) QCD penguin operators}
\begin{eqnarray}
{\renewcommand\arraystretch{1.5}
\begin{array}{ll}
\displaystyle
O_3\, =\, (\bar{d}_\alpha b_\alpha)_{V-A}\sum_{q}(\bar{q}_\beta q_\beta)_{V-A}\;,
& \displaystyle
O_4\, =\, (\bar{d}_\alpha b_\beta)_{V-A}\sum_{q}(\bar{q}_\beta q_\alpha)_{V-A}\;,
\\
\displaystyle
O_5\, =\, (\bar{d}_\alpha b_\alpha)_{V-A}\sum_{q}(\bar{q}_\beta q_\beta)_{V+A}\;,
& \displaystyle
O_6\, =\, (\bar{d}_\alpha b_\beta)_{V-A}\sum_{q}(\bar{q}_\beta q_\alpha)_{V+A}\;;
\end{array}}
\label{eq:operators-2}
\end{eqnarray}

\item[]{(3) electroweak penguin operators}
\begin{eqnarray}
{\renewcommand\arraystretch{1.5}
\begin{array}{ll}
\displaystyle
O_7\, =\,
\frac{3}{2}(\bar{d}_\alpha b_\alpha)_{V-A}\sum_{q}e_{q}(\bar{q}_\beta q_\beta)_{V+A}\;,
& \displaystyle
O_8\, =\,
\frac{3}{2}(\bar{d}_\alpha b_\beta)_{V-A}\sum_{q}e_{q}(\bar{q}_\beta q_\alpha)_{V+A}\;,
\\
\displaystyle
O_9\, =\,
\frac{3}{2}(\bar{d}_\alpha b_\alpha)_{V-A}\sum_{q}e_{q}(\bar{q}_\beta q_\beta)_{V-A}\;,
& \displaystyle
O_{10}\, =\,
\frac{3}{2}(\bar{d}_\alpha b_\beta)_{V-A}\sum_{q}e_{q}(\bar{q}_\beta q_\alpha)_{V-A}\;.
\end{array}}
\label{eq:operators-3}
\end{eqnarray}
\end{enumerate}
Here $\alpha$ and $\beta$ are $SU(3)$ color indices. Then the calculation of decay amplitude is to compute the hadronic matrix
elements of the local operators.


\begin{figure}[!!htb]
\centering
\begin{tabular}{l}
\includegraphics[width=0.8\textwidth]{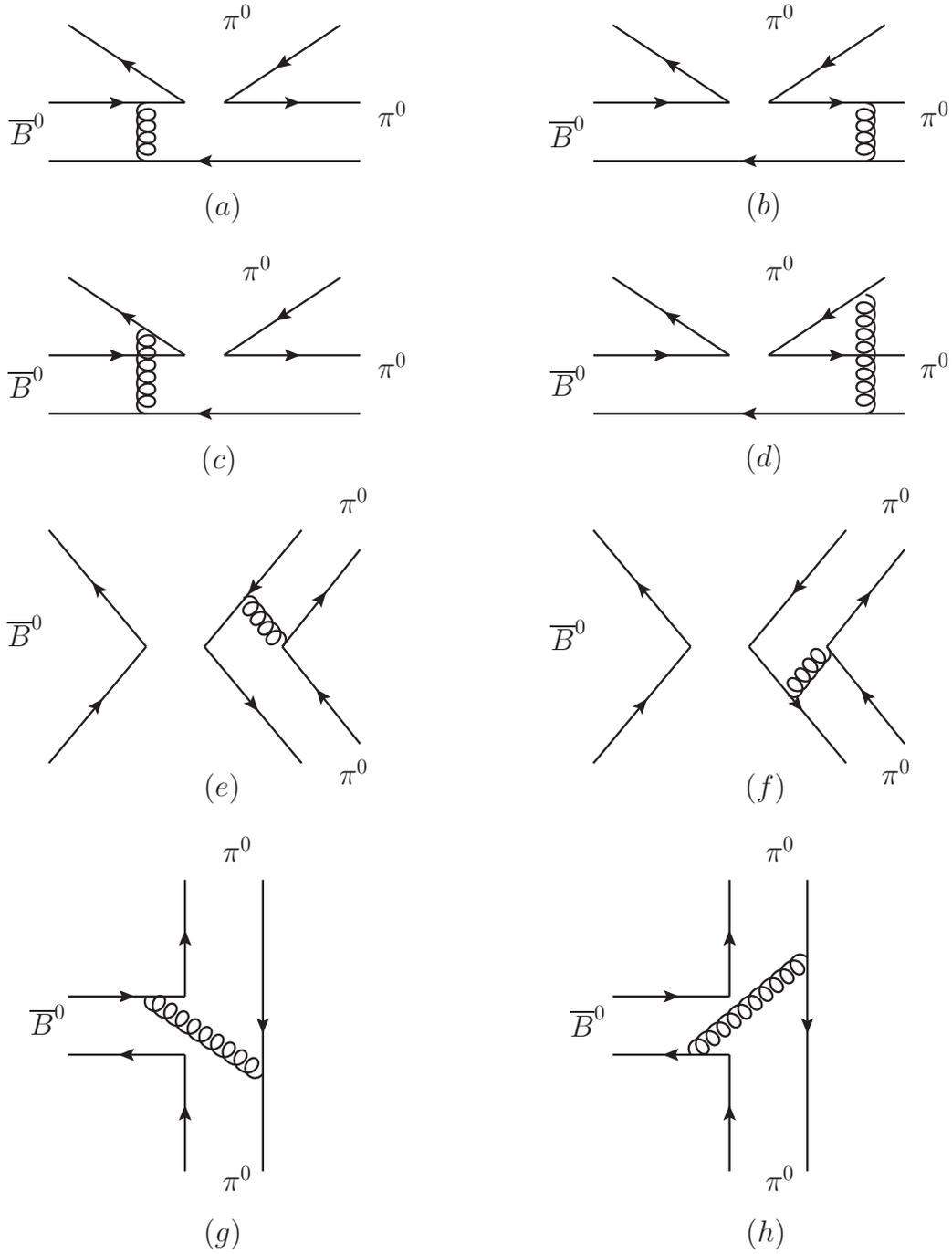}
\end{tabular}
\caption{Typical Feynman diagrams contributing to the
$\bar{B}^{0}\to\pi^{0}\pi^{0}$ decays in the PQCD approach
at leading order.}
  \label{fig:fig1}
\end{figure}

In the PQCD, the soft ($\Phi$), hard ($H$), and harder ($C$) dynamics characterized by different scales
make up the decay amplitude.
It is conceptually written as follows:
\beq
Amplitude\sim\int d^{4} k_{1} d^{4} k_{2} d^{4} k_{3} Tr[C(t)\Phi_{\bar{B}^{0}}(k_{1}) \Phi_{\pi^{0}}(k_{2})
\Phi_{\pi^{0}}(k_{3})H(k_{1}, k_{2}, k_{3}, t)],
\label{eq:exp6}
\eeq
where $k_{i}$ are the momenta of light quarks included in each meson,
and $Tr$ denotes the trace over Dirac and color indices.
The Wilson coefficient $C(t)$ results from the radiative corrections at short distance. The non-perturbative part
is absorbed into wave function $\Phi_{M}$, which is universal and channel independent.
$H$ describes the four quark operator and the quark pair produced by
 a gluon whose scale is at the order of $M_{B}$, so this hard part $H$ can be perturbative calculated.

We consider the $B$ meson at rest for simplicity and assume that
the light final states poin meson moving along the direction of $n=(1,0,0_{T})$ and
$v=(0,1,0_{T})$. It is convenient to use the light-cone coordinate $(P^{+},P^{-},P_{T})$ to describe the meson's momenta,
where, $P^{\pm}=\frac{1}{\sqrt{2}}(p^{0}\pm p^{3}), P_{T}=(p^{1},p^{2})$.
Working at the rest frame of $\bar{B}^{0}$ meson, the momenta of
$\bar{B}^{0}$, $\pi^{0}$, and $\pi^{0}$ can be written as follows:
\beq
&P_{1} = {M_{B}\over\sqrt{2}} (1,1,0_{T})\ \non
&P_{2} = {M_{B}\over\sqrt{2}} (0,1,0_{T})\ \non
&P_{3} = {M_{B}\over\sqrt{2}} (1,0,0_{T})
\label{eq:exp7}
\eeq

Putting the light (anti-) quark momenta in $\bar{B}^{0}$, $\pi^{0}$ and $\pi^{0}$
as $k_{1}, k_{2}, k_{3}$, respectively, we can choose:
\beq
&k_{1} = (x_{1}p^{+}_{1},0,k_{1T})\ \non
&k_{2} = (0,x_{2}p^{-}_{2},k_{2T})\ \non
&k_{3} = (x_{3}p^{+}_{3},0,k_{3T})
\label{eq:exp8}
\eeq
Then, integrating over $k^{+}_{1}, k^{-}_{2}, k^{+}_{3}$ in Eq.~(\ref{eq:exp6}) leads to
\beq
\begin{split}
Amplitude\sim\int &d^{4} x_{1} d^{4} x_{2} d^{4} x_{3} b_{1} db_{1} b_{2} db_{2} b_{3} db_{3}\\
&\times Tr[C(t) \Phi_{\bar{B}^{0}}(x_{1}, b_{1}) \Phi_{\pi^{0}}(x_{2}, b_{2})
\Phi_{\pi^{0}}(x_{3}, b_{3})H(x_{i}, b_{i}, t)]e^{-S(t)},
\label{eq:exp9}
\end{split}
\eeq
where $b_{i}$ is the conjugate space coordinate of $k_{iT}$, and $t$ the largest energy scale in $H$.
The exponential Sudakov factor $e^{-S(t)}$ comes
from higher order radiative corrections to wave functions and hard amplitudes, it suppresses the soft dynamics effectively~\cite{liTseng1998}
and thus make a reliable perturbative calculation of the hard part $H$.

Fig.~\ref{fig:fig1} shows the lowest order diagrams to be calculated in the PQCD approach for $\bar{B}^{0}\to\pi^{0}\pi^{0}$ decay.
The sum contributions of the non-factorizable diagrams $(a)$ and $(b)$ which come from the operator $O_{2}$ are
\beq
\begin{split}
{\cal M}_{a} =& \frac{-1}{\sqrt{2N_{c}}}32 \pi C_{F}M_{B}^{2}
\int_{0}^{1} dx_{1} dx_{2} dx_{3} \int_{0}^{\infty} b_{1} db_{1} b_{2} db_{2} \Phi_{B}(x_{1}, b_{1})
\{[(x_{2}-2) \Phi_{\pi}^{A}(x_{2})\Phi_{\pi}^{A}(x_{3}) \\& + r_{\pi} (1-2x_{2}) \Phi_{\pi}^{T}(x_{2})
\Phi_{\pi}^{A}(x_{3})+r_{\pi} (1-2x_{2}) \Phi_{\pi}^{P}(x_{2})
\Phi_{\pi}^{A}(x_{3}) ] \alpha_{s}(t_{a}^{1}) h_{a}^{1}(x_{1}, x_{2}, x_{3}, b_{1},b_{2})\\&
\exp[-S_{B}(t_{a}^{1}) -S_{\pi}(t_{a}^{1}) -S_{\pi}(t_{a}^{1})] C(t_{a}^{1}) - 2r_{\pi} \Phi_{\pi}^{P}(x_{2})
\Phi_{\pi}^{A}(x_{3}) \alpha_{s}(t_{a}^{2})\\& h_{a}^{2}(x_{1}, x_{2}, x_{3}, b_{1}, b_{2})
\exp[-S_{B}(t_{a}^{2}) -S_{\pi}(t_{a}^{2}) -S_{\pi}(t_{a}^{2})] C(t_{a}^{2})\},
\label{eq:exp10}
\end{split}
\eeq
where $C_{F}=4/3$ is the group factor of the $SU(3)_{c}$ gauge group and $r_{\pi}=M_{0\pi}/M_{B}$.
 The wave function $\Phi_{M}$, the functions $h_{a}^{1,2}(x_{1}, x_{2}, x_{3}, b_{1}, b_{2})$,
and the Sudakov factor $S_{X}(t_{i})(X=\bar{B}^{0}, \pi^{0}, \pi^{0})$ will be given in the appendix.

The total contribution for the non-factorizable diagrams $(c)$ and $(d)$ is
\beq
\begin{split}
{\cal M}_{c} =& \frac{-1}{\sqrt{2N_{c}}} 32\pi C_{F}M_{B}^{2}
\int_{0}^{1} dx_{1} dx_{2} dx_{3}\int_{0}^{\infty} b_{2} db_{2} b_{3} db_{3} \Phi_{B}(x_{1},b_{3}) \{[\Phi_{\pi}^{A}(x_{2})\Phi_{\pi}^{A}(x_{3})(1-x_{1}-x_{3})\\& +r_{\pi}\Phi_{\pi}^{P}(x_{2})
\Phi_{\pi}^{A}(x_{3})(1-x_{2})+r_{\pi}\Phi_{\pi}^{T}(x_{2})\Phi_{\pi}^{A}(x_{3})(1-x_{2})]
\alpha_{s}(t_{c}^{1})h_{c}^{1}(x_{1},x_{2},x_{3},b_{2},b_{3})\\
&\exp[-S_{B}(t_{c}^{1}) -S_{\pi}(t_{c}^{1}) -S_{\pi}(t_{c}^{1})]
C(t_{c}^{1})+[-\Phi_{\pi}^{A}(x_{2})\Phi_{\pi}^{A}(x_{3})(1+x_{3}-x_{1}-x_{2})
-r_{\pi}\Phi_{\pi}^{P}(x_{2})\Phi_{\pi}^{A}(x_{3})(1-x_{2})\\&+r_{\pi}\Phi_{\pi}^{T}(x_{2})
\Phi_{\pi}^{A}(x_{3})(1-x_{2})]\alpha_{s}(t_{c}^{2})
h_{c}^{2}(x_{1},x_{2},x_{3},b_{2},b_{3})\exp[-S_{B}(t_{c}^{2}) -S_{\pi}(t_{c}^{2}) -S_{\pi}(t_{c}^{2})]
C(t_{c}^{2})\}.
\label{eq:exp11}
\end{split}
\eeq

The factorizable annihilation diagrams $(e)$ and $(f)$ which come from the operators $O_{1}, O_{3}, O_{4}, O_{5},O_{6}, O_{7}, O_{8}, O_{9}, O_{10}$
 involve only two light mesons wave functions. $M_{e}$ is for $(V-A)(V-A)$ and $(V-A)(V+A)$ type operators, and $M_{e}^{p}$ is for $(1+\gamma_{5})(1-\gamma_{5})$ type operators:
\beq
\begin{split}
{\cal M}_{e} =&  8 S \pi C_{F}M_{B}^{2}\int_{0}^{1} dx_{2} dx_{3}
\int_{0}^{\infty} b_{2} db_{2} b_{3} db_{3}\{[-\Phi_{\pi}^{A}(x_{2})\Phi_{\pi}^{A}(x_{3})
x_{2}-2r_{\pi}^{2}\Phi_{\pi}^{P}(x_{2})\Phi_{\pi}^{P}(x_{3})(1+x_{2})\\& +2r_{\pi}^{2}
\Phi_{\pi}^{T}(x_{2})\Phi_{\pi}^{P}(x_{3})(x_{2}-1)] \alpha_{s}(t_{e}^{1})h_{e}^{1}(x_{2},x_{3},b_{2},b_{3})
 \exp[ -S_{\pi}(t_{e}^{1}) -S_{\pi}(t_{e}^{1})]C(t_{e}^{1})\\& +[\Phi_{\pi}^{A}(x_{2})\Phi_{\pi}^{A}(x_{3})
x_{3} +2r_{\pi}^{2}\Phi_{\pi}^{P}(x_{2})\Phi_{\pi}^{P}(x_{3})(1+x_{3}) +2r_{\pi}^{2}
\Phi_{\pi}^{P}(x_{2})\Phi_{\pi}^{T}(x_{3})(1-x_{3})]\\&\alpha_{s}(t_{e}^{2})h_{e}^{2}(x_{2},x_{3},b_{2},b_{3})
\exp[ -S_{\pi}(t_{e}^{2}) -S_{\pi}(t_{e}^{2})]C(t_{e}^{2})\},
\label{eq:exp12}
\end{split}
\eeq
\beq
\begin{split}
{\cal M}_{e}^{P} =&  8 S  \pi C_{F} M_{B}^{2}\int_{0}^{1} dx_{2} dx_{3}
\int_{0}^{\infty} b_{2} db_{2} b_{3} db_{3}\{[-r_{\pi}\Phi_{\pi}^{P}(x_{2})\Phi_{\pi}^{A}(x_{3})
x_{2}-r_{\pi}\Phi_{\pi}^{T}(x_{2})\Phi_{\pi}^{A}(x_{3})x_{2}\\&-2r_{\pi}\Phi_{\pi}^{A}(x_{2})\Phi_{\pi}^{P}(x_{3})]
\alpha_{s}(t_{e}^{1})h_{e}^{1}(x_{2},x_{3},b_{2},b_{3})\exp[ -S_{\pi}(t_{e}^{1}) -S_{\pi}(t_{e}^{1})]C(t_{e}^{1})
+[-2r_{\pi}\Phi_{\pi}^{P}(x_{2})\Phi_{\pi}^{A}(x_{3})\\&-r_{\pi}\Phi_{\pi}^{A}(x_{2})\Phi_{\pi}^{P}(x_{3})x_{3}
-r_{\pi}\Phi_{\pi}^{A}(x_{2})\Phi_{\pi}^{T}(x_{3})x_{3}]
\alpha_{s}(t_{e}^{2})h_{e}^{2}(x_{2},x_{3},b_{2},b_{3})
\exp[ -S_{\pi}(t_{e}^{2}) -S_{\pi}(t_{e}^{2})]C(t_{e}^{2})\},
\label{eq:exp13}
\end{split}
\eeq
where $S=2$ comes from the requirement of identity principle. The non-factorizable annihilation diagrams $(g)$ and $(h)$ come from the operators $O_{4}, O_{6}, O_{8}, O_{10}$.  $M_{g}$ is the
contribution containing the operator of type $(V-A)(V-A)$, and $M_{g}^{P}$ is the
contribution containing the operator of type $(1+\gamma_{5})(1-\gamma_{5})$.

\beq
\begin{split}
{\cal M}_{g} =& \frac{1}{\sqrt{2N_{c}}}32 S \pi C_{F}M_{B}^{2}
\int_{0}^{1} dx_{1} dx_{2} dx_{3} \int_{0}^{\infty} b_{1} db_{1} b_{2} db_{2} \Phi_{B}(x_{1}, b_{1})
 \{[(x_{1}+x_{3}) \Phi_{\pi}^{A}(x_{2}) \Phi_{\pi}^{A}(x_{3})\\& + r_{\pi}^{2} (2+x_{1}+x_{2}+x_{3})
\Phi_{\pi}^{P}(x_{2})\Phi_{\pi}^{P}(x_{3}) -r_{\pi}^{2}\Phi_{\pi}^{P}(x_{2})\Phi_{\pi}^{T}(x_{3})
(x_{2}-x_{1}-x_{3})+r_{\pi}^{2}\Phi_{\pi}^{T}(x_{2})\Phi_{\pi}^{P}(x_{3})(x_{1}+x_{3}-x_{2})\\
 & -r_{\pi}^{2}\Phi_{\pi}^{T}(x_{2})\Phi_{\pi}^{T}(x_{3})(2-x_{1}-x_{2}-x_{3})]
 \alpha_{s}(t_{g}^{1}) h_{g}^{1}(x_{1}, x_{2}, x_{3}, b_{1},b_{2})
\exp[-S_{B}(t_{g}^{1}) -S_{\pi}(t_{g}^{1}) -S_{\pi}(t_{g}^{1})] C(t_{g}^{1})\\&+[-\Phi_{\pi}^{A}(x_{2})
\Phi_{\pi}^{A}(x_{3})x_{2} +r_{\pi}^{2}\Phi_{\pi}^{P}(x_{2})\Phi_{\pi}^{P}(x_{3})(x_{1}-x_{2}-x_{3})
-r_{\pi}^{2}\Phi_{\pi}^{P}(x_{2})\Phi_{\pi}^{T}(x_{3})(x_{1}-x_{3}+x_{2})\\& -r_{\pi}^{2}
\Phi_{\pi}^{T}(x_{2})\Phi_{\pi}^{P}(x_{3})(x_{1}-x_{3}+x_{2}) +r_{\pi}^{2}
\Phi_{\pi}^{T}(x_{2})\Phi_{\pi}^{T}(x_{3})(x_{1}-x_{2}-x_{3})]
 \alpha_{s}(t_{g}^{2})\\& h_{g}^{2}(x_{1}, x_{2}, x_{3}, b_{1}, b_{2})
\exp[-S_{B}(t_{g}^{2}) -S_{\pi}(t_{g}^{2}) -S_{\pi}(t_{g}^{2})] C(t_{g}^{2})\},
\label{eq:exp14}
\end{split}
\eeq
\beq
\begin{split}
{\cal M}_{g}^{P} =& \frac{-1}{\sqrt{2N_{c}}} 32 S \pi C_{F}M_{B}^{2}
\int_{0}^{1} dx_{1} dx_{2} dx_{3} \int_{0}^{\infty} b_{1} db_{1} b_{2} db_{2} \Phi_{B}(x_{1}, b_{1})
\{[-\Phi_{\pi}^{A}(x_{2}) \Phi_{\pi}^{A}(x_{3})x_{2}\\& -r_{\pi}^{2}(2+x_{1}+x_{2}+x_{3})
\Phi_{\pi}^{P}(x_{2})\Phi_{\pi}^{P}(x_{3})-r_{\pi}^{2}\Phi_{\pi}^{P}(x_{2})\Phi_{\pi}^{T}(x_{3})(x_{2}-x_{1}-x_{3})\\
& +r_{\pi}^{2}\Phi_{\pi}^{T}(x_{2})\Phi_{\pi}^{P}(x_{3})(x_{1}+x_{3}-x_{2}) +r_{\pi}^{2}
\Phi_{\pi}^{T}(x_{2})\Phi_{\pi}^{T}(x_{3})(x_{1}+x_{2}+x_{3}-2)]\alpha_{s}(t_{g}^{1})
h_{g}^{1}(x_{1}, x_{2}, x_{3}, b_{1},b_{2})\\& \exp[-S_{B}(t_{g}^{1}) -S_{\pi}(t_{g}^{1}) -S_{\pi}(t_{g}^{1})]
C(t_{g}^{1}) +[-\Phi_{\pi}^{A}(x_{2})\Phi_{\pi}^{A}(x_{3})(x_{1}-x_{3})-r_{\pi}^{2}
\Phi_{\pi}^{P}(x_{2})\Phi_{\pi}^{P}(x_{3})(x_{1}-x_{2}-x_{3})\\& -r_{\pi}^{2}
\Phi_{\pi}^{P}(x_{2})\Phi_{\pi}^{T}(x_{3})(x_{1}-x_{3}+x_{2}) +r_{\pi}^{2}
\Phi_{\pi}^{T}(x_{2})\Phi_{\pi}^{P}(x_{3})(x_{1}+x_{2}-x_{3})-r_{\pi}^{2}
\Phi_{\pi}^{T}(x_{2})\Phi_{\pi}^{T}(x_{3})(x_{2}+x_{3}-x_{1})]\\
& \alpha_{s}(t_{g}^{2}) h_{g}^{2}(x_{1}, x_{2}, x_{3}, b_{1}, b_{2})
\exp[-S_{B}(t_{g}^{2}) -S_{\pi}(t_{g}^{2}) -S_{\pi}(t_{g}^{2})] C(t_{g}^{2})\},
\label{eq:exp15}
\end{split}
\eeq

The total decay amplitude of $\bar{B}^{0}\to\pi^{0}\pi^{0}$ is then
\beq
\begin{split}
{\cal \bar{A}}(\bar{B}^{0}\to\pi^{0}\pi^{0}) =&V^*_{ud}V_{ub}[C_{1}{\cal M}_{\rm e}f_{B}+C_{2}({\cal M}_{\rm a}+{\cal M}_{\rm c})]-
V^*_{td}V_{tb}[(2C_{3}+\frac{5}{3}C_{4}+2C_{5}+\frac{2}{3}C_{6}+\frac{1}{2}C_{7}\\&+\frac{1}{6}C_{8}
+\frac{1}{2}C_{9}-\frac{1}{3}C_{10}){\cal M}_{\rm e}f_{B}+(C_{6}-\frac{1}{2}C_{8}){\cal M}_{\rm e}^{\rm P}f_{B}+(2C_{4}
+\frac{1}{2}C_{10}){\cal M}_{\rm g}+(2C_{6}+\frac{1}{2}C_{8}){\cal M}_{\rm g}^{\rm P}]
\label{eq:exp16}
\end{split}
\eeq
and the decay width is expressed as
\beq
\Gamma(\bar{B}^{0}\to\pi^{0}\pi^{0})= \frac{G_{F}^{2}M_{B}^{3}}{128\pi}|{\cal  \bar{A}}(\bar{B}^{0}\to\pi^{0}\pi^{0})|^{2}
\label{eq:exp17}
\eeq

The decay amplitude of the charge conjugate channel for $\bar{B}^{0}\to\pi^{0}\pi^{0}$ can be obtained by replacing
$V^*_{ud}V_{ub}$ to $V_{ud}V^{*}_{ub}$ and $V^*_{td}V_{tb}$ to $V_{td}V^{*}_{tb}$ in Eq.~(\ref{eq:exp16}).
The decay amplitude of $\bar{B}^{0}\to\pi^{0}\pi^{0}$ in Eq.~(\ref{eq:exp16}) can be parameterized as
\beq
{\cal \bar{A}}&=& V^*_{ud}V_{ub}T - V^*_{td}V_{tb}P
=V^*_{ud}V_{ub}T[1+ze^{i(-\alpha+\delta)}],
\label{eq:exp18}
\eeq
where $z=|V^*_{td}V_{tb}/V^*_{ud}V_{ub}||P/T|$, and $\delta=\arg(P/T)$ is the relative strong phase between tree diagrams $T$ and
penguin diagrams $P$. $z$ and $\delta$
can be calculated from PQCD.

Similarly, the decay amplitude for $B^{0}\to\pi^{0}\pi^{0}$ can be parameterized as
\beq
{\cal A} &=& V^*_{ub}V_{ud}T - V^*_{tb}V_{td}P
=V^*_{ub}V_{ud}T[1+ze^{i(\alpha+\delta)}].
\label{eq:exp19}
\eeq

\section{Numerical Evaluation and discussions of results} \label{sec:numer}

The parameters have been used in numerical calculation~\cite{Beringer2012,Amhis,J.Charles:2005,A.Hocher:2001,H.-n.Li:2005} are shown in Table~\ref{tab:Vapa}.

\begin{table}[htpb]
\caption{The values of parameters adopted in numerical evaluation.}
\label{tab:Vapa}
\begin{center} \vspace{-0.3cm}{\small
\begin{tabular}[t]{c|c|c|c|c|c|c|c|c|c|c|c}
\hline  \hline
 parameters  & $\Lambda_{QCD}^{f=4}$  &   $m_{W}$
             &$m_{B}$                 &   $f_{\pi}$
             &$f_{B}$                 &   $m_{0\pi}$
             &$\tau_{B^0}$            &   $|V^*_{ud}V_{ub}|$
             &$|V^*_{td}V_{tb}|$      &   \\
\hline
values        &$\hspace{0.3cm}0.25$ {\rm GeV}    &   \hspace{0.3cm}$80.41$ {\rm GeV}
              &$\hspace{0.3cm}5.280$ {\rm GeV}   &   $\hspace{0.3cm}0.13$ {\rm GeV}
              &$\hspace{0.3cm}0.21$ {\rm GeV}    &   $\hspace{0.3cm}1.4$ {\rm GeV}
              &$\hspace{0.3cm} 1.55\times10^{-12}$ {\rm s}
              &$\hspace{0.3cm} 0.00346$
              &$\hspace{0.3cm}0.00885$ &      \\
\hline \hline
\end{tabular}}
\end{center}
\end{table}

\begin{figure}[!!htb]
\begin{center}
\includegraphics[width=0.6 \textwidth]{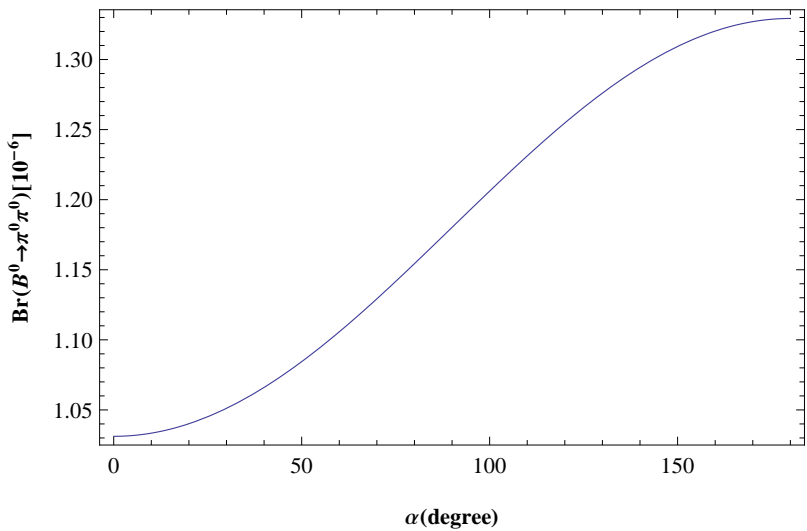}
\end{center}
\caption{The averaged branching ratio  of
    $\bar{B}^{0} (B^{0})\to \pi^{0}\pi^{0}$ decay as a function of CKM angle $\alpha$.}
\label{fig:fig2}
\end{figure}

We leave the Cabibbo-Kobayashi-Maskawa (CKM) phase angle $\alpha$ as a free parameter to
explore the branching ratio and CP asymmetry. From Eqs.~(\ref{eq:exp18}) and (\ref{eq:exp19}), we get the averaged decay width for $\bar{B}^{0}(B^{0})\to\pi^{0}\pi^{0}$
\beq
\Gamma(\bar{B}^{0}(B^{0})\to\pi^{0}\pi^{0}) &=&\frac{G_{F}^{2}M_{B}^{3}}{128\pi}
(\frac{|{\cal A}|^{2}}{2}+\frac{|{\cal \bar{A}}|^{2}}{2})
\nonumber\\
&=& \frac{G_{F}^{2}M_{B}^{3}}{128\pi}|V^*_{ud}V_{ub}T|^{2}[1+2z\cos(\alpha)\cos(\delta)+z^{2}].
\label{eq:exp20}
\eeq
Using the above parameters, we get $z=0.52$ and $\delta=106^{\circ}$ in PQCD. Equation~(\ref{eq:exp20}) is a function of CKM angle $\alpha$.
In Fig.~\ref{fig:fig2}, we plot the averaged branching ratio of the decay $\bar{B}^{0}(B^{0})\to\pi^{0}\pi^{0}$ with respect
to the parameter $\alpha$. Since the CKM angle $\alpha$ is constrained as $\alpha$
around $85^{\circ}$~\cite{A.Hocher:2001}.
\beq
\alpha=(85.4_{-3.8}^{+3.9})^{\circ}
\label{eq:exp22}
\eeq

We can arrive from Fig.~\ref{fig:fig2}
\beq
1.15 \times 10^{-6} <Br(\bar{B}^{0}(B^{0})\to\pi^{0}\pi^{0}) < 1.18 \times 10^{-6},  \hspace{0.5cm}
for 80^{\circ} <\alpha < 90^{\circ}
\label{eq:exp23}
\eeq

The number $z=|V^*_{td}V_{tb}/V^*_{ud}V_{ub}||P/T|= 0.52$ means that the amplitude of
 penguin diagrams is about 0.52 times that of tree diagrams, which shows though the tree contribution dominate this decay, the penguin contribution
 cannot be ignored, i. e., there are large contributions from both tree diagrams and penguin diagrams.

 Besides the phase angle $\alpha$, the major theoretical errors come from the uncertainties of $\omega_{b}=0.4\pm0.04$ GeV, $f_{B}=0.21\pm0.02$ GeV,
 and the Gegenbauer moment $a_{2}^{\pi}=0.25\pm0.15$. Taking into account the uncertainties of these parameters, we find
\beq
Br(\bar{B}^{0} (B^{0})\to \pi^{0}\pi^{0})=\left[1.17_{-0.08}^{+0.09}(\omega_{b})_{-0.07}^{+0.05}(f_{B})_{-0.06}^{+0.02}(a_{2}^{\pi})\right]\times 10^{-6}.
\label{eq:exbe}
\eeq
When all important theoretical errors from different sources, including those from the uncertainty of phase angle $\alpha$, are added in quadrature, we get $Br(\bar{B}^{0} (B^{0})\to \pi^{0}\pi^{0})=(1.17_{-0.12}^{+0.11})\times 10^{-6}$.

 In the literature, there already exist a lot of studies on $B^{0}\rightarrow \pi^{0}\pi^{0}$ decay. We give some recent works devoted to
the resolution of the challenge:

(a) In Ref.~\cite{QChang2014}, Qin Chang and Junfeng Sun $et$  $al$ do a global fit on the spectator
scattering and annihilation parameters$ X_{H}(\rho_{H}, \phi_{H})$,
$X_{A}^{i}(\rho_{A}^{i}, \phi_{A}^{i})$
and $X_{A}^{f}(\rho_{A}^{f}, \phi_{A}^{f})$ for the available experimental data for $B_{u,d}\rightarrow \pi\pi,\pi K$ and $K\bar{K}$ decays
 in the QCDF framework. They obtained large ${B}^{0}\to\pi^{0}\pi^{0}$ branching ratios $(1.67_{-0.30}^{+0.33})\times 10^{-6}$ and
$(2.13_{-0.38}^{+0.43})\times 10^{-6}$ for different scenarios.

(b)In Ref.~\cite{XLiu2015}, Xin Liu , Hsiang-nan Li and Zhen-Jun Xiao investigate the Glauber-gluon effect on the $B\rightarrow \pi\pi$ and
$\rho\rho$ decays based on
the $k_{T}$ factorization theorem, they observed significant modification of $B^{0}\rightarrow \pi^{0}\pi^{0}$ branching ratio through a transverse-momentum-dependent(TMD) wave function for the pion with a weak falloff in parton transverse
momentum $k_{T}$. They get the branching ratio of $B^{0}\rightarrow \pi^{0}\pi^{0}$ $0.61\times 10^{-6}$.

(c)In Ref.~\cite{CFQiao2015}, Cong-Feng Qiao $et$  $al$ give a possible solution to the $B\rightarrow\pi\pi$ puzzle using the
Principle of Maximum Conformality(PMC). They applied the PMC procedure to the QCDF analysis with the goal of eliminating the renormalization
scale ambiguity and achieving an accurate pQCD prediction which is independent of theoretical conventions. They found the pQCD prediction is highly sensitive to the choice of the renormalization scale which
enter the decay amplitude, they obtained $Br(B_{d}\rightarrow\pi^{0}\pi^{0})=(0.98^{+0.44}_{-0.31})\times10^{-6}$ by applying the principle
of maximum conformality. However, we find the PQCD prediction is not sensitive to the choice of the renormalization scale for this decay based on our calculation. In our approach, we set the renormalization scale $\mu=t$(the largest energy scale in $H$) to diminish the large logarithmic radiative corrections and minimize the NLO contributions to the form factors. By changing the hard scale $t$ from $0.9t$ to $1.3t$, we find the branching ratio of $B^{0}\rightarrow \pi^{0}\pi^{0}$ change a little. The choice of the renormalization scale is not a main reason for the $B^{0}\rightarrow \pi^{0}\pi^{0}$ puzzle, even when the NLO
contributions are taken into account~\cite{Ya-Lan Zhang:2015}.

(d)In Ref.~\cite{Ya-Lan Zhang:2015}, Ya-Lan Zhang $et$  $al$ performed a systematic study for the $B\rightarrow (\pi^{+}\pi^{-},\pi^{+}\pi^{0},\pi^{0}\pi^{0})$ decays in the PQCD factorization approach with the inclusion of all currently known NLO contributions from various sources. They got the NLO PQCD prediction for $B^{0}\rightarrow \pi^{0}\pi^{0}$ branching ratio $Br(B^{0}\rightarrow \pi^{0}\pi^{0})=[0.23_{-0.05}^{+0.08}(\omega(b))_{-0.04}^{+0.05}(f_{B})_{-0.03}^{+0.04}(a_{2}^{\pi})]\times 10^{-6}$, it is still much smaller than
the measured data.

(e)In Ref.~\cite{H.Y.Cheng:2015}, Hai-Yang Cheng, Cheng-Wei Chiang and An-Li Kuo  used
flavor SU(3) symmetry to analyze the data of charmless $B$ meson decays to two pseudoscalar mesons $(PP)$ and one
vector and one pseudoscalar mesons $(VP)$ . They found the color-suppressed tree amplitude larger than previously known and
has a strong phase of $-70^{\circ}$ relative to the color favored tree amplitude in the PP sector, this large color-suppressed tree amplitude results in the
large $B^{0}\rightarrow \pi^{0}\pi^{0}$ branching ratios $1.43\pm0.55\times 10^{-6}$ and
$1.88\pm 0.42\times 10^{-6}$ for different scheme.

\begin{table}[htbp]
\centering
\caption{The pQCD predictions for the CP-averaged branching ratios(in unit of $10^{-6}$).}{\centering}
\label{tab:brnp}
\begin{tabular}[t]{c|c|c|c|c|c|c|c|c|c|c|c}
\hline \hline
Channel        &LO~\cite{LUY} &NLO~\cite{H.-n.Li:2005} &NLO~\cite{Ya-Lan Zhang:2015}&LO(this work) &QCDF~\cite{Beneke2003} & BABAR Data~\cite{Heavy} & Belle Data~\cite{Heavy} & HFAG Data ~\cite{Heavy}& \\
\hline
$B^{0}\rightarrow \pi^{0}\pi^{0}$ &$0.12$  &$0.29$ &$0.23$ &$1.17_{-0.12}^{+0.11}$ &$0.3$ &$1.83\pm 0.21 \pm 0.13$ &$0.90\pm 0.12 \pm 0.10$ &$1.17\pm 0.13$&\\
\hline  \hline
\end{tabular}
\end{table}

There are some works on $B^{0}\rightarrow \pi^{0}\pi^{0}$ decay in the framework of PQCD approach before\cite{LUY,H.-n.Li:2005,Ya-Lan Zhang:2015}, we list these
numerical values in Table~\ref{tab:brnp}. Ref.~\cite{LUY} is the earliest PQCD calculations for $B^{0}\rightarrow \pi^{0}\pi^{0}$ decay at the leading order(LO), Hsiang-nan Li $et$  $al$ considered partial NLO contributions in Ref.~\cite{H.-n.Li:2005}. Based on the work of Refs.~\cite{LUY,H.-n.Li:2005}, Ya-Lan Zhang $et$  $al$ calculated all currently known NLO contributions from various sources in Ref.~\cite{Ya-Lan Zhang:2015}. As shown in Table~\ref{tab:brnp}, one can see that
the NLO contributions are much larger than LO contributions for $B^{0}\rightarrow \pi^{0}\pi^{0}$ decay in previous works. Despite this, it is still much smaller than the experimental data. In this work, we recalculate the $B^{0}\rightarrow \pi^{0}\pi^{0}$ decay in the framework of PQCD approach at LO. Our result is much larger than that of previous predictions\cite{LUY,H.-n.Li:2005,Ya-Lan Zhang:2015}, there are two reasons that make the difference. For the operator $O_1 =
(\bar{u}_\alpha u_\alpha)_{V-A}(\bar{d}_\beta b_\beta)_{V-A}$, it can contribute not only to non-factorizable diagrams (a) and (b), but to factorizable annihilation diagrams (e) and (f)(see Fig.~\ref{fig:fig1}) as well. We find the largest contributions come from the factorizable annihilation diagrams $(e)$ and $(f)$, which
come from tree operator $O_{1}$ and penguin operators $O_{3}, O_{4}, O_{5},O_{6}, O_{7}, O_{8}, O_{9}, O_{10}$. In previous PQCD works\cite{LUY,H.-n.Li:2005,Ya-Lan Zhang:2015}, first, the contributions of the factorizable annihilation diagrams $(e)$ and $(f)$ come from tree operator $O_{1}$ had not been taken into account, the authors only considered the non-factorizable diagrams $(a)$ and $(b)$(small contributions) for operator $O_{1}$; second, For $O_{3}, O_{4}, O_{9}, O_{10}$ operators, previous calculations\cite{LUY} showed their contributions cancel between diagrams $(e)$ and $(f)$, however, we recalculate it and find their contributions cannot be canceled between diagrams $(e)$ and $(f)$, as shown in Eqs.(\ref{eq:exp12})(\ref{eq:exp13}). If we get rid of the contributions of ${\cal M}_{\rm e}$ and ${\cal M}_{\rm e}^{\rm P}$ terms in Eq.~(\ref{eq:exp16}), our result is $Br(\bar{B}^{0}(B^{0})\to\pi^{0}\pi^{0})\sim 0.26\times10^{-6}$, which is consistent with previous PQCD predictions\cite{LUY,H.-n.Li:2005,Ya-Lan Zhang:2015}.
The hard scale $t$ in Eq.~(\ref{eq:exp9}) characterizes the size of NLO contributions, by changing the hard scale $t$ from $0.9t$ to $1.3t$,
we find the branching ratio of $B^{0}\rightarrow \pi^{0}\pi^{0}$ changes about $~10\%$, which means although the NLO diagrams may make a significant contributions to $B^{0}\rightarrow \pi^{0}\pi^{0}$ decay\cite{H.-n.Li:2005,Ya-Lan Zhang:2015}, the LO contributions still dominate this decay. Because
there are identical particles in final state for this decay, one must consider identical principle. Usually the decay width receives a
symmetry factor $1/2$ due to the identical particles in the final state, but in our calculations, we have calculated the symmetrized Feynman diagrams and all these contributions have been included in the total decay amplitude formula(\ref{eq:exp16}), and hence there is no need to add an extra factor in decay width.
In our recalculations, we consider all the possible diagrams's contribution, including non-factorizable contributions and annihilation contributions. We obtain the branching ratio of $B^{0}\to\pi^{0}\pi^{0}$ $(1.17_{-0.12}^{+0.11})\times10^{-6}$, which is still smaller than BABAR result~\cite{Heavy}, but it is consistent with the Belle and HFAG results~\cite{Heavy}. More experimental and theoretical efforts should be made to resolve the $B^{0}\rightarrow \pi^{0}\pi^{0}$ puzzle.

\begin{figure}[htbp!]
\begin{center}
\includegraphics[width=0.6 \textwidth]{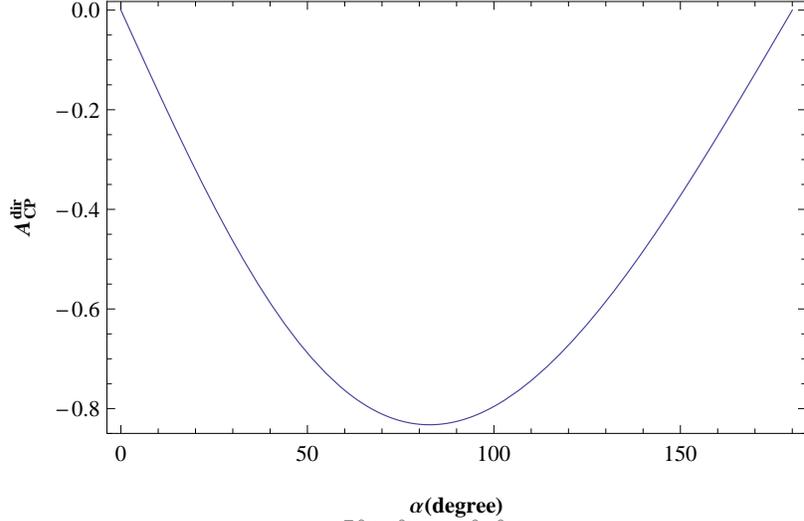}
\vspace{-0.5cm} \caption{Direct CP violation parameter of
$\bar{B}^{0} (B^{0})\to \pi^{0}\pi^{0}$ decay as a function of CKM angle
$\alpha$.}\label{fig:fig3}
\end{center}
\end{figure}
\par

In SM , the CKM phase angle is the origin of CP violation. Using Eqs.(\ref{eq:exp18}) and (\ref{eq:exp19}), the direct CP violating parameter is
\beq
\acp^{\rm dir} &=& \frac{|\cal\bar{A}|^{\rm 2}-|\cal A|^{\rm 2}}{|\cal\bar{A}|^{\rm 2}+|\cal A|^{\rm 2}}
=\frac{2z\sin(\alpha)\sin(\delta)}{1+2z\cos(\alpha)\cos(\delta)+z^{2}}
\label{eq:exp24}
\eeq
It is approximately proportional to CKM angle $\sin(\alpha)$, strong phase $\sin(\delta)$, and the relative size $z$ between
penguin contribution and tree contribution. We show the direct CP asymmetry $\acp^{\rm dir}$ in Fig.~\ref{fig:fig3}.
One can see from this figure that the direct CP asymmetry parameter of $\bar{B}^{0}(B^{0})\to\pi^{0}\pi^{0}$
can be as large as from $-83\%$ to $-82\%$ when $80^{\circ}<\alpha<90^{\circ}$. The large direct CP asymmetry is also a result
of there are large contributions from both tree diagrams and penguin diagrams in this decays.

For the neutral $B^{0}$ decays, the $\bar{B}^{0}- B^{0}$ mixing is very complex.
Following notations in the previous literature ~\cite{G.Kramer:1997}, we define the mixing induced CP
violation parameter as
\beq
a_{\epsilon+\epsilon'} = \frac{-2Im(\lambda_{CP})}{1+|\lambda_{CP}|^{2}},
\label{eq:exp25}
\eeq
where
\beq
\lambda_{CP} = \frac{V^*_{tb}V_{td}<\pi^{0}\pi^{0}|H_{eff}|\bar{B}^{0}>}{V_{tb}V^{*}_{td}<\pi^{0}\pi^{0}|H_{eff}|B^{0}>}.
\label{eq:exp26}
\eeq
Using equations (\ref{eq:exp18}) and (\ref{eq:exp19}), we can derive as
\beq
\lambda_{CP} = e^{2i\alpha}\frac{1+ze^{i(\delta-\alpha)}}{1+ze^{i(\delta+\alpha)}}
\label{eq:exp27}
\eeq

\begin{figure}[htbp!]
\begin{center}
\includegraphics[width=0.6 \textwidth]{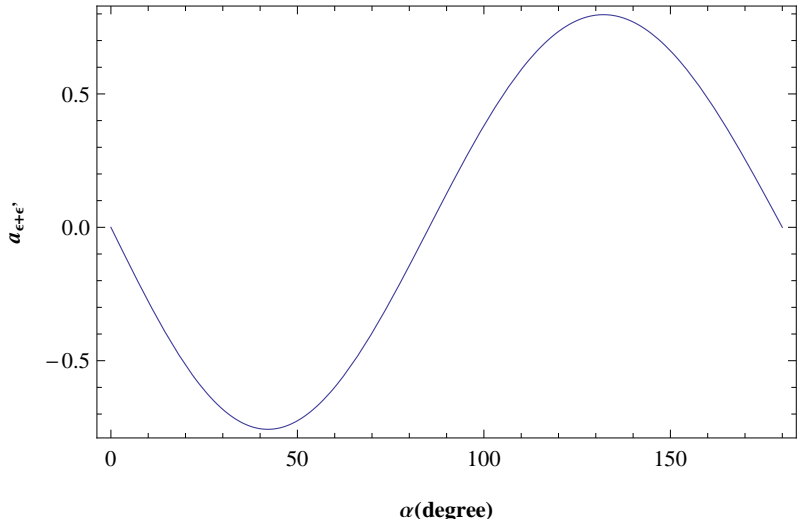}
\vspace{-0.5cm} \caption{Mixing CP violation parameter of
$\bar{B}^{0} (B^{0})\to \pi^{0}\pi^{0}$ decay as a function of CKM angle
$\alpha$.}\label{fig:fig4}
\end{center}
\end{figure}
\par

 If $z$ is a very small number, i. e., the penguin diagram contribution is suppressed comparing with the tree diagram contribution, the
mixing induced CP asymmetry parameter $a_{\epsilon+\epsilon'}$ is proportional to $-\sin2\alpha$, which will be a good place for the CKM
angle $\alpha$ measurement. However as we have already mentioned, $z$ is not very small. We give the mixing CP asymmetry in Fig.~\ref{fig:fig4}, one
can see that $a_{\epsilon+\epsilon'}$ is not a simple  $-\sin2\alpha$ behavior because of the so-called penguin pollution. It is close to $6\%$ when the angle near $85^{\circ}$. At present, there are
no CP asymmetry measurements in experiment but the possible large CP
violation we predict for $\bar{B}^{0} (B^{0})\to \pi^{0}\pi^{0}$ decays  might be observed in the coming Belle-II experiments.


\section{Summary} \label{sec:summary}

In this work, we recalculate the branching ratio and CP asymmetries of the decays $\bar{B}^{0} (B^{0})\to \pi^{0}\pi^{0}$
in PQCD approach at LO. From our calculations, we find the branching ratio of $B^{0}\to\pi^{0}\pi^{0}$ $(1.17_{-0.12}^{+0.11})\times10^{-6}$, much larger
 than that of previous predictions\cite{LUY}, and there are large CP violation in this process, which may be measured in the coming Belle-II experiments. The branching ratio we get is still smaller than BABAR
 result~\cite{Heavy}, but it is consistent with the latest Belle and HFAG
results~\cite{Heavy}.

\begin{acknowledgments}

The authors thank
Dr. Ming-Zhen Zhou and Dr. Wen-Long Sang for valuable discussions.
This work is supported by the National Natural Science Foundation of
 China under Grant Nos.11047028 and 11645002, and by the Fundamental
 Research Funds of the Central Universities under Grant Number
 XDJK2012C040.

\end{acknowledgments}

\section{Appendix : Formulae For The Calculations Used In The Text} \label{sec:appendix}

We present the explicit expressions of the formulae used in Sec.~\ref{sec:pert} in the appendix.
The expressions of the meson distribution amplitudes $\Phi_{M}$ are given at first.
For $B^{0}$ meson wave function, we use the function~\cite{LUY,Keum2001,Y.-Y.Keum:2001}
\beq
\phi_{B}(x,b)&=& N_Bx^2(1-x)^2
\exp\left[-\frac{1}{2}\left(\frac{xm_B}{\omega_b}\right)^2
-\frac{\omega_b^2 b^2}{2}\right] \;,
\label{eq:propagator1}
\eeq
The parameter $\omega_{b}= 0.4$ GeV is constrained by other charmless $B$ decays~\cite{LUY,Keum2001,Y.-Y.Keum:2001}.
For the $\pi$ meson's wave function, the distribution amplitude $\Phi_{\pi}^{A}$ for the twist-2 wave function and
the distribution amplitude $\Phi_{\pi}^{P}$ and $\Phi_{\pi}^{T}$ of the twist-3 wave
functions are taken from  ~\cite{H.-n.Li:2005,Z.J.Xiao:2008,Y.Y.Fan:2013,Z.J.Xiao:2012}
\begin{eqnarray}
\Phi_{\pi}^{A}(x) &=& \frac{3f_{\pi}}{\sqrt{2N_c}} x(1-x) \times [1+a_{1}^{\pi}C_{1}^{\frac{3}{2}}(2x-1)
+a_{2}^{\pi}C_{2}^{\frac{3}{2}}(2x-1) +a_{4}^{\pi}C_{4}^{\frac{3}{2}}(2x-1)]; \nonumber\\
\Phi_{\pi}^{P}(x)&=&\frac{f_{\pi}}{2\sqrt{2N_{c}}}\left[1+(30\eta_{3}-\frac{5}{2}\rho_{\pi}^{2})C_{2}^{\frac{1}{2}}(2x-1)
-3\{\eta_{3}\omega_{3}+\frac{9}{20}\rho_{\pi}^{2}(1+6a_{2}^{\pi})\}C_{4}^{\frac{1}{2}}(2x-1)\right] ;\label{eq:propagator2}\nonumber\\
\Phi_{\pi}^{T}(x) &=& \frac{f_{\pi}}{2\sqrt{2N_{c}}}(1-2x)\left[1+6\left(5\eta_{3}-\frac{1}{2}\eta_{3}\omega_{3}-\frac{7}{20}\rho_{\pi}^{2}
-\frac{3}{5}\rho_{\pi}^{2}a_{2}^{\pi}\right)(1-10x+10x^{2})\right],
\end{eqnarray}
where $a_{i}^{\pi}$ are the Gegenbauer moments, the mass ratio $\rho_{\pi}=m_{\pi}/m_{0\pi}$. The Gegenbauer polynomials are defined by ~\cite{H.-n.Li:2005}.
\begin{eqnarray}\label{eq:propagator3}
C_{2}^{\frac{1}{2}}(t) &=& \frac{1}{2}(3t^{2}-1); \nonumber\\
C_{4}^{\frac{1}{2}}(t) &=& \frac{1}{8}(35t^{4}-30t^{2}+3); \nonumber\\
C_{2}^{\frac{3}{2}}(t) &=& \frac{3}{2}(5t^{2}-1);\nonumber\\
C_{4}^{\frac{3}{2}}(t) &=& \frac{15}{8}(21t^{4}-14t^{2}+1); \nonumber\\
C_{1}^{\frac{3}{2}}(t) &=& 3t,
\end{eqnarray}
 and the Gegenbauer moments and other parameters are adopted from Refs.\cite{H.-n.Li:2005,P.Ball:2005}
\begin{align}
 a_{1}^{\pi}=0, && \quad a_{2}^{\pi}=0.25, && \quad  a_{4}^{\pi}=-0.015, \nonumber\\
  \rho_{\pi}=m_{\pi}/m_{0\pi}, &&\quad \eta_{3}=0.015,   && \quad  \omega_{3}=-3.0
\end{align}
with $m_{0\pi}$ the chiral mass of the pion.

$S_{\bar{B}^{0}}$, $S_{\pi^{0}}$, $S_{\pi^{0}}$ used in the decay amplitudes are defined as
\beq
S_{\bar{B}^{0}}(t) &=& s(x_{1}P_{1}^{+},b_{1})+ 2\int_{\frac{1}{b_{1}}}^{t}
\frac{d\bar{\mu}}{\bar{\mu}}\gamma(\alpha_{s}(\bar{\mu})),
\label{eq:propagator4}
\eeq
\beq
S_{\pi^{0}}(t) &=& s(x_{2}P_{2}^{-},b_{2})+ s((1-x_{2})P_{2}^{-},b_{2})+2\int_{\frac{1}{b_{2}}}^{t}
\frac{d\bar{\mu}}{\bar{\mu}}\gamma(\alpha_{s}(\bar{\mu})),
\label{eq:propagator5}
\eeq
\beq
S_{\pi^{0}}(t) &=& s(x_{3}P_{3}^{+},b_{3})+s((1-x_{3})P_{3}^{+},b_{3})
+ 2\int_{\frac{1}{b_{3}}}^{t}
\frac{d\bar{\mu}}{\bar{\mu}}\gamma(\alpha_{s}(\bar{\mu})),
\label{eq:propagator6}
\eeq
where the so called Sudakov factor $s(Q,b)$ resulting from the resummation
of double logarithms is given as ~\cite{H.-n.Li:2003, H.-n.Li:1999}
\beq
s(Q,b) = \int_{\frac{1}{b}}^{Q}\frac{d\mu}{\mu}\left[\ln(\frac{Q}{\mu})
A(\alpha(\bar{\mu}))+B(\alpha_{s}(\bar{\mu}))\right]
\label{eq:propagator7}
\eeq
with
\beq
A &=& C_{F}\frac{\alpha_{s}}{\pi}+\left[\frac{67}{9}-\frac{\pi^{2}}{3}-\frac{10}{27}n_{f}
+ \frac{2}{3}\beta_{0}\ln(\frac{e^{\gamma_{E}}}{2})\right](\frac{\alpha_{s}}{\pi})^{2},
\label{eq:propagator8}
\eeq
\beq
B &=& \frac{2}{3} \frac{\alpha_{s}}{\pi} \ln\left(\frac{e^{2\gamma_{E}-1}}{2}\right),
\label{eq:propagator9}
\eeq
here $\gamma_{E}=0.57722\cdot\cdot\cdot$ is the Euler constant, $n_{f}$ is the
active quark flavor number.

The functions $h_{i}(i=a,c,e,g)$ come from the Fourier transformation
of propagators of virtual quark and gloun in the hard part calculations.
They are given as follow

\begin{align}
& h_{a}^{j}(x_{1},x_{2},x_{3},b_{1},b_{2})= \nonumber \\
&
\biggl\{\theta(b_{1}-b_{2})I_{0}(M_{B} \sqrt{x_{1}(1-x_{2})}b_{2})K_{0}(M_{B} \sqrt{x_{1}(1-x_{2})}b_{1})
\nonumber \\
& \qquad\qquad\qquad\qquad + (b_1\leftrightarrow b_2) \biggr\}
\times\left(
\begin{matrix}
 \mathrm (K_{0}(M_{B}F_{a(j)}b_{1}), & \text{for}\quad F^2_{a(j)}>0 \\
 \frac{\pi i}{2} \mathrm{H}_0^{(1)}(M_{B} \sqrt{|F_{a(j)}^{2}|}b_{1}), &
 \text{for}\quad F^2_{a(j)}<0
 \end{matrix}\right),
\label{eq:propagator10}
\end{align}
where $F_{a(j)}$'s are defined by

\beq
F_{a(1)}^{2} = 1-x_{2}\ , \non
F_{a(2)}^{2} =x_{1} \;. \label{eq:propagator11}
\eeq
\begin{align}
& h_{c}^{j}(x_{1},x_{2},x_{3},b_{2},b_{3}) =  \nonumber \\
&
\biggl\{\theta(b_{2}-b_{3})I_{0}(M_{B} \sqrt{x_{1}(1-x_{2})}b_{3})K_{0}(M_{B} \sqrt{x_{1}(1-x_{2})}b_{2})
\nonumber \\
& \qquad\qquad\qquad\qquad + (b_2\leftrightarrow b_3) \biggr\}
\times\left(
\begin{matrix}
\mathrm (K_{0}(M_{B}F_{c(j)}b_{3}), & \text{for}\quad F^2_{c(j)}>0 \\
 \frac{\pi i}{2} \mathrm{H}_0^{(1)}(M_{B} \sqrt{|F_{c(j)}^{2}|}b_{3}), &
 \text{for}\quad F^2_{c(j)}<0
 \end{matrix}\right),
\label{eq:propagator12}
\end{align}
where $F_{c(j)}$'s are defined by
\beq
&F_{c(1)}^{2} = x_{1}+x_{2}+x_{3}-x_{1}x_{2}-x_{2}x_{3}-1\ , \non
&F_{c(2)}^{2} =x_{1}-x_{3}-x_{1}x_{2}+x_{2}x_{3}\;. \label{eq:propagator11}
\eeq

\beq
h_{e}^{1}(x_{2},x_{3},b_{2},b_{3}) =&S_{t}(x_2) K_{0}(M_{B}\sqrt{x_{2}x_{3}}b_{3})
\nonumber \\
 & \times \{\theta(b_{2}-b_{3})I_{0}(M_{B}\sqrt{x_{2}}b_{2})
 K_{0}(M_{B}\sqrt{x_{2}}b_{3})+(b_{2}\leftrightarrow b_{3})\},
\label{eq:propagator14}
\eeq
\beq
h_{e}^{2}(x_{2},x_{3},b_{2},b_{3}) =&S_{t}(x_3) K_{0}(M_{B}\sqrt{x_{2}x_{3}}b_{2})
\nonumber \\
&\times \{\theta(b_{2}-b_{3})I_{0}(M_{B}\sqrt{x_{3}}b_{3})
 K_{0}(M_{B}\sqrt{x_{3}}b_{2})+(b_{2}\leftrightarrow b_{3})\}.
\label{eq:propagator15}
\eeq

\begin{align}
& h_{g}^{j}(x_{1},x_{2},x_{3},b_{1},b_{2}) =  \nonumber \\
&
\biggl\{\theta(b_{1}-b_{2})I_{0}(M_{B} \sqrt{x_{2}x_{3}}b_{1})K_{0}(M_{B} \sqrt{x_{2}x_{3}}b_2)
\nonumber \\
& \qquad\qquad\qquad\qquad + (b_1\leftrightarrow b_2) \biggr\}
\times\left(
\begin{matrix}
\mathrm (K_{0}(M_{B}F_{g(j)}b_{1}), & \text{for}\quad F^2_{g(j)}>0 \\
 \frac{\pi i}{2} \mathrm{H}_0^{(1)}(M_{B} \sqrt{|F_{g(j)}^{2}|}b_{1}), &
 \text{for}\quad F^2_{g(j)}<0
 \end{matrix}\right),
\label{eq:propagator16}
\end{align}

where $F_{g(j)}$'s are defined by
\beq
&F_{g(1)}^{2} = x_{1}+x_{2}+x_{3}-x_{1}x_{2}-x_{2}x_{3}\ , \non
&F_{g(2)}^{2} = x_{1}x_{2}-x_{2}x_{3}\;.\label{eq:propagator17}
\eeq

We adopt the parametrization for $S_{t}(x)$ contributing to
the factorizable diagrams ~\cite{T.Kurimoto:2003}

\beq
S_{t}(x) = \frac{2^{1+2c}\Gamma(\frac{3}{2}+c)}{\sqrt{\pi}\Gamma(1+c)}
 [x(1-x)]^{c}
\label{eq:propagator18}
\eeq
where the parameter $c=0.3$.
The hard scale $t$ in the amplitudes are taken as the largest energy scale in $H$ to kill the large logarithmic radiative corrections:
\beq
\begin{split}
& t_{a}^{1}=\max(M_{B}\sqrt{\mid F_{a(1)}^{2}\mid},M_{B}\sqrt{x_{1}(1-x_{2})},\frac{1}{b_{1}},
 \frac{1}{b_{2}})\;; \non\\
& t_{a}^{2}=\max(M_{B}\sqrt{\mid F_{a(2)}^{2}\mid},M_{B}\sqrt{x_{1}(1-x_{2})},\frac{1}{b_{1}},
 \frac{1}{b_{2}})\;; \non\\
& t_{c}^{1}=\max(M_{B}\sqrt{\mid F_{c(1)}^{2}\mid},M_{B}\sqrt{x_{1}(1-x_{2})},\frac{1}{b_{2}},
 \frac{1}{b_{3}})\;; \non\\
& t_{c}^{2}=\max(M_{B}\sqrt{\mid F_{c(2)}^{2}\mid},M_{B}\sqrt{x_{1}(1-x_{2})},\frac{1}{b_{2}},
 \frac{1}{b_{3}})\;; \non\\
& t_{e}^{1}=\max(M_{B}\sqrt{x_{2}},\frac{1}{b_{2}},\frac{1}{b_{3}})\;; \non\\
& t_{e}^{2}=\max(M_{B}\sqrt{x_{3}},\frac{1}{b_{2}},\frac{1}{b_{3}})\;; \non\\
& t_{g}^{1}=\max(M_{B}\sqrt{\mid F_{g(1)}^{2}\mid},M_{B}\sqrt{x_{2}x_{3}},\frac{1}{b_{1}},
 \frac{1}{b_{2}})\;; \label{eq:propagator19}\\
& t_{g}^{2}=\max(M_{B}\sqrt{\mid F_{g(2)}^{2}}\mid,M_{B}\sqrt{x_{2}x_{3}},\frac{1}{b_{1}},
 \frac{1}{b_{2}}).
\label{eq:propagator19}
\end{split}
\eeq

\end{document}